\documentclass[
reprint, amsmath, amssymb, aps,
citeautoscript,
prb, floatfix]{revtex4-2}
\usepackage[colorlinks=false]{hyperref}
\usepackage{float}
\usepackage{graphicx}%
\usepackage{dcolumn}%
\usepackage{bm}%
\usepackage{xcolor}
\usepackage[normalem]{ulem}
\usepackage{color,soul}
\usepackage{float} 
\usepackage[]{lineno}
\begin{document}

\preprint{APS/123-QED}

\title{Single-Mode Squeezed Light Generation and Tomography with an Integrated Optical Parametric Oscillator}

\author{Taewon~Park\textsuperscript{1,2}}\author{Hubert~S.~Stokowski\textsuperscript{1}}
\author{Vahid~Ansari\textsuperscript{1}}
\author{Samuel~Gyger\textsuperscript{1}}
\author{Kevin~K.~S.~Multani\textsuperscript{1}}
\author{Oguz~Tolga~Celik\textsuperscript{1,2}}
\author{Alexander~Y.~Hwang\textsuperscript{1}}\author{Devin~J.~Dean\textsuperscript{1}}
\author{Felix~M.~Mayor\textsuperscript{1}}
\author{Timothy~P.~McKenna\textsuperscript{3}}

\author{Martin~M.~Fejer\textsuperscript{1}}
\author{Amir~H.~Safavi-Naeini\textsuperscript{1,{$\star$}}}

\affiliation{%
\textsuperscript{1} \mbox{Department of Applied Physics and Ginzton Laboratory, Stanford University, Stanford, California 94305, USA}
}%
\affiliation{%
\textsuperscript{2} \mbox{Department of Electrical Engineering, Stanford University, Stanford, California 94305, USA}
}%
\affiliation{%
\textsuperscript{3} \mbox{Physics \& Informatics Laboratories, NTT Research, Inc., Sunnyvale, California 94085, USA}
}%
\affiliation{%
\textsuperscript{{$\star$} } safavi@stanford.edu
}%


\pacs{Valid PACS appear here}
\maketitle


{\bf
Quantum optical technologies promise advances in sensing, computing, and communication. A key resource is squeezed light, where quantum noise is redistributed between optical quadratures. We introduce a monolithic, chip-scale platform that exploits the $\chi^{(2)}$ nonlinearity of a thin-film lithium niobate (TFLN) resonator device to efficiently generate squeezed states of light.  Our system integrates all essential components---except for the laser and two detectors---on a single chip with an area of one square centimeter, significantly reducing the size, operational complexity, and power consumption associated with conventional setups. Our work addresses challenges that have limited previous integrated nonlinear photonic implementations that rely on either $\chi^{(3)}$ nonlinear resonators or on  integrated waveguide $\chi^{(2)}$ parametric amplifiers. Using the balanced homodyne measurement subsystem that we implemented on the same chip, we measure a squeezing of 0.55 dB and an anti-squeezing of 1.55 dB. We use 20 mW of input power to generate the parametric oscillator pump field by employing second harmonic generation on the same chip. Our work represents a substantial step toward compact and efficient quantum optical systems posed to leverage the rapid advances in integrated nonlinear and quantum photonics. }

\renewcommand{\figurename}{Fig}
\begin{figure*}[t]
  \begin{center}
      \includegraphics[width=\textwidth]{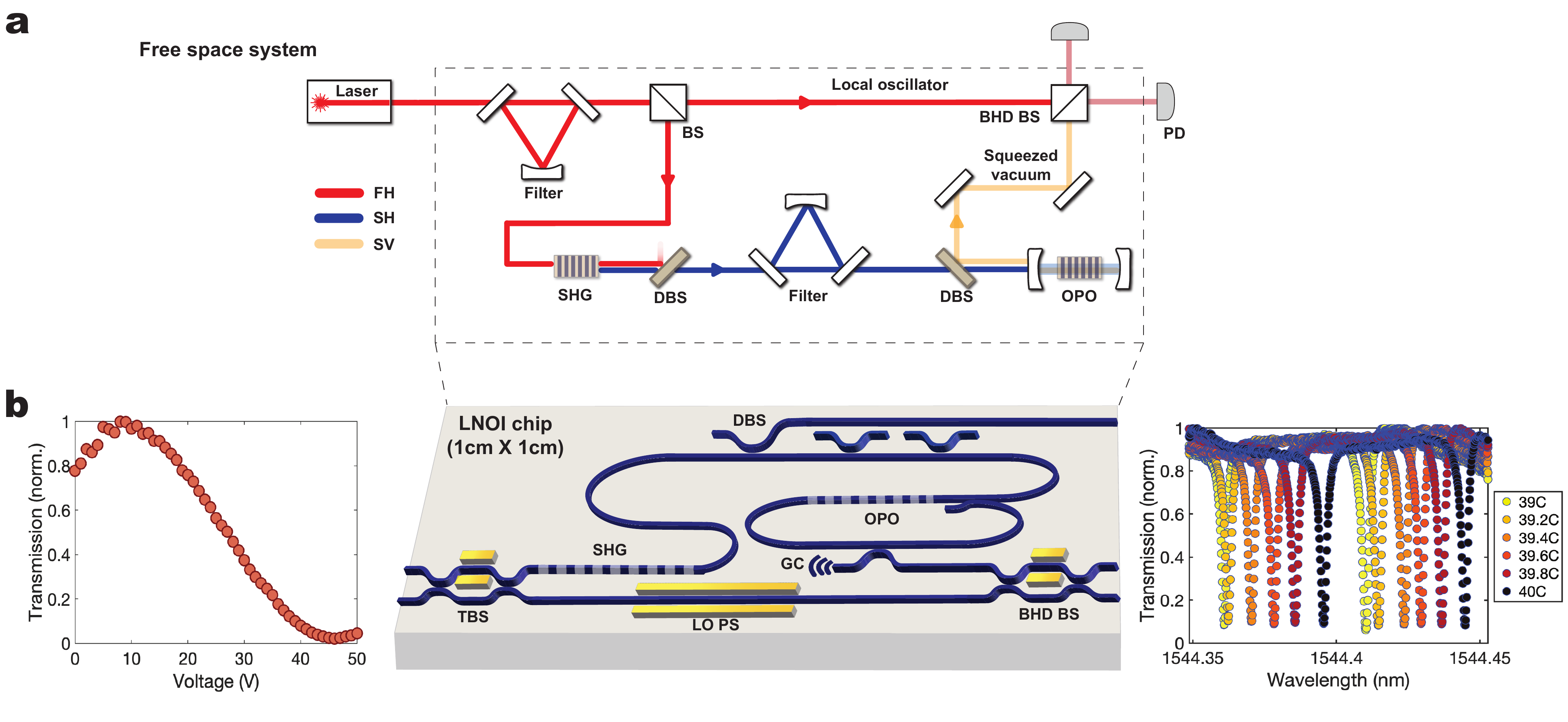}
  \end{center}
 \caption{\textbf{Free space system and a monolithic quantum photonic integrated circuit for generating and analyzing squeezed light using a sub-threshold optical parametric oscillator.}
\textbf{(a)} Free space system for generating and analyzing squeezed vacuum state of light from a sub-threshold optical parametric oscillator (OPO). Output of the laser at the fundamental harmonic (FH) goes through a filter to get a clean mode profile and gets split into two paths using a beamsplitter (BS). One path goes to the second harmonic generator (SHG) section to prepare pump light for the OPO while the light on the other path acts as a local oscillator (LO). The remaining FH light after SHG is filtered out using a dichroic beamsplitter (DBS). Second harmonic light goes through a filter for clean mode profile and pumps the OPO below oscillation threshold. The output from the OPO at the FH is a squeezed vacuum state of light. The generated squeezed light is sent to the 50:50 BS (BHD BS) and superimposed with the LO. The two outputs that are balanced in intensity are sent to the photodiodes. The fluctuations of the difference between the two photocurrents from the photodiodes are measured using a radio frequency spectrum analyzer (RFSA). \textbf{(b)}  A monolithic quantum photonic integrated circuit (1 cm$^2$ in area) on a periodically poled thin-film lithium niobate on insulator (LNOI) platform. Dotted box in (a) illustrates the components integrated into a single photonic chip. The photonic integrated circuit consists of directional couplers, tunable beamsplitter (TBS), waveguide SHG, DBS, OPO, grating coupler (GC), local oscillator phase shifter (LO PS), and BHD BS. The left plot shows the tuning behavior of the TBS with applied DC voltage where $V_{\pi}\sim$ 35 V (see Supplementary Information). The right plot shows the tuning behavior of the mode spectrum of the OPO cavity at different thermoelectric cooler (TEC) temperature settings. The mode spectrum shifts by the entire FSR ($\sim$50 pm) when we change the temperature setting of the TEC by 1.2\textdegree C.}
 \label{fig:fig1}
\end{figure*}

In the drive to improve the precision and performance of sensing systems, quantum optical phenomena are increasingly becoming indispensable \cite{pelucchi2022potential, polino2020photonic}. The essential resources in this respect are the squeezed states of light, which are quantum states that have reduced noise along one of their quadrature components \cite{walls1983squeezed}. The development of integrated photonics, enabling the miniaturization of optical devices and systems on a chip scale, offers a natural avenue to further the capabilities of quantum optical systems and engineer them to address a broader range of real sensing problems \cite{moody20222022}. However, creating a portable and efficient squeezed light source has remained a formidable challenge \cite{bongs2023quantum}.

The generation of squeezed states of light can employ resonant or nonresonant schemes, as well as either $\chi^{(2)}$ or $\chi^{(3)}$ optical nonlinearities~\cite{yurke1984use,schnabel2017squeezed}. In bulk optics, resonant approaches with $\chi^{(2)}$ nonlinearities have been extensively demonstrated~\cite{wu1986generation, schnabel2017squeezed}, achieved remarkable levels of squeezing,~\cite{vahlbruch2016detection} and been deployed in real quantum sensors~\cite{aasi2013enhanced}. Resonant schemes for generating squeezing, \emph{i.e.}, the use of optical parametric oscillators (OPOs), offer several advantages: these methods squeeze the optical field with less pump power, they provide natural mode-matching, and they also enhance responsivity to signals in a way that can potentially combine with squeezing~\cite{peano2015intracavity}. 

Advances in the field of nanophotonics have enabled the integration of OPOs on chip \cite{levy2010cmos, lu2021ultralow, mckenna2022ultra}. The burgeoning interest and developments in quantum computing \cite{paesani2019generation,vaidya2020broadband,arrazola2021quantum} have fueled the development of integrated OPOs for generating squeezed light. These demonstrations have used the $\chi^{(3)}$ nonlinearity to successfully generate squeezed light in \emph{nondegenerate} mode, where the signal and idler are at different optical frequencies. For many quantum sensing and continuous variable (CV) quantum computing applications, \emph{degenerate} operation of a squeezer is highly desirable. The use of the $\chi^{(3)}$ makes degenerate squeezing challenging as the nearly resonant and considerable circulating power of the pump field either requires operation at cryogenic temperatures ~\cite{safavi2013squeezed} or the use of more complex cavity designs and pumping schemes to mitigate noise processes~\cite{zhao2020near, zhang2021squeezed} that arise.

Our work demonstrates degenerate squeezed light generation and tomography by implementing an integrated photonic circuit incorporating an OPO. Our device works as a resonant $\chi^{(2)}$ squeezer, and we measure squeezing of 0.55 dB and anti-squeezing of 1.55 dB with 20 mW of input fundamental harmonic power. We demonstrate a monolithic chip that incorporates all essential components—bar the laser and two detectors—streamlining the system by reducing operational complexity, shrinking the size, reducing power consumption, and addressing environmental constraints. 

Our approach uses a periodically poled thin film lithium niobate (TFLN) \cite{zhu2021integrated,boes2023lithium} and leverages its strong $\chi^{(2)}$ nonlinearity, electro-optic effect, and low-loss linear optical characteristics for the photonic circuit. Integrated TFLN devices have shown remarkable promise, offering a toolbox of capabilities such as efficient light generation spanning from mid-infrared to near-UV \cite{wang2018ultrahigh, lu2020toward, mishra2021mid, park2022high, hwang2023tunable}, dispersion engineering \cite{jankowski2020ultrabroadband, jankowski2021dispersion, mishra2022ultra}, electro-optic modulation and tuning \cite{wang2018integrated, celik2022high}, optical parametric oscillation \cite{lu2021ultralow, mckenna2022ultra}, waveguide squeezers \cite{nehra2022few, stokowski2023integrated}, and frequency combs \cite{zhang2019broadband, stokowski2023FMOPO}. Here, we harness the full capabilities of this emerging platform, demonstrating a monolithic integrated quantum system for the generation and tomography of squeezed light.

Our device is a photonic integrated chip, with a surface area of roughly 1 cm\textsuperscript{2}, that houses  nearly all the required components to generate and analyze the squeezed vacuum state of light (Fig. 1(b)). We fabricated the chip on TFLN (see Supplementary Information). We couple fundamental harmonic (FH) light in the telecom C-band from a continuous wave (CW) laser into the chip and then send it to a tunable beamsplitter (TBS). The TBS splits the light into two paths: one leads to the waveguide second harmonic generation (SHG) section to prepare the pump at the second harmonic frequency for the OPO, the other serves as a local oscillator (LO). We adjust the division of light into these paths by tuning the DC voltage applied to the TBS electrodes. For light entering the SHG path, after propagating through the SHG section and generating frequency-doubled SH light, we filter the FH light out using three identical on-chip dichroic beam splitters (DBS). The SH light is then sent into another nonlinear section within a resonator and is used to pump an OPO to induce parametric gain in the cavity. This pump field only makes a single pass through the cavity due to the presence of the wavelength-dependent output coupler. As a result, below the oscillation threshold, a squeezed vacuum state at the fundamental harmonic emerges. For a balanced homodyne measurement, we combine the squeezed vacuum state and the prepared LO through the balanced homodyne beamsplitter (BHD BS). We adjust the DC voltage on the BHD BS electrode to balance the intensities of both outputs. By changing the DC voltage on the LO phase shifter, we can sweep the LO phase and measure the variance of the squeezed vacuum state across various reference phases.
\begin{figure}
  \begin{center}
      \includegraphics[width=1\columnwidth]{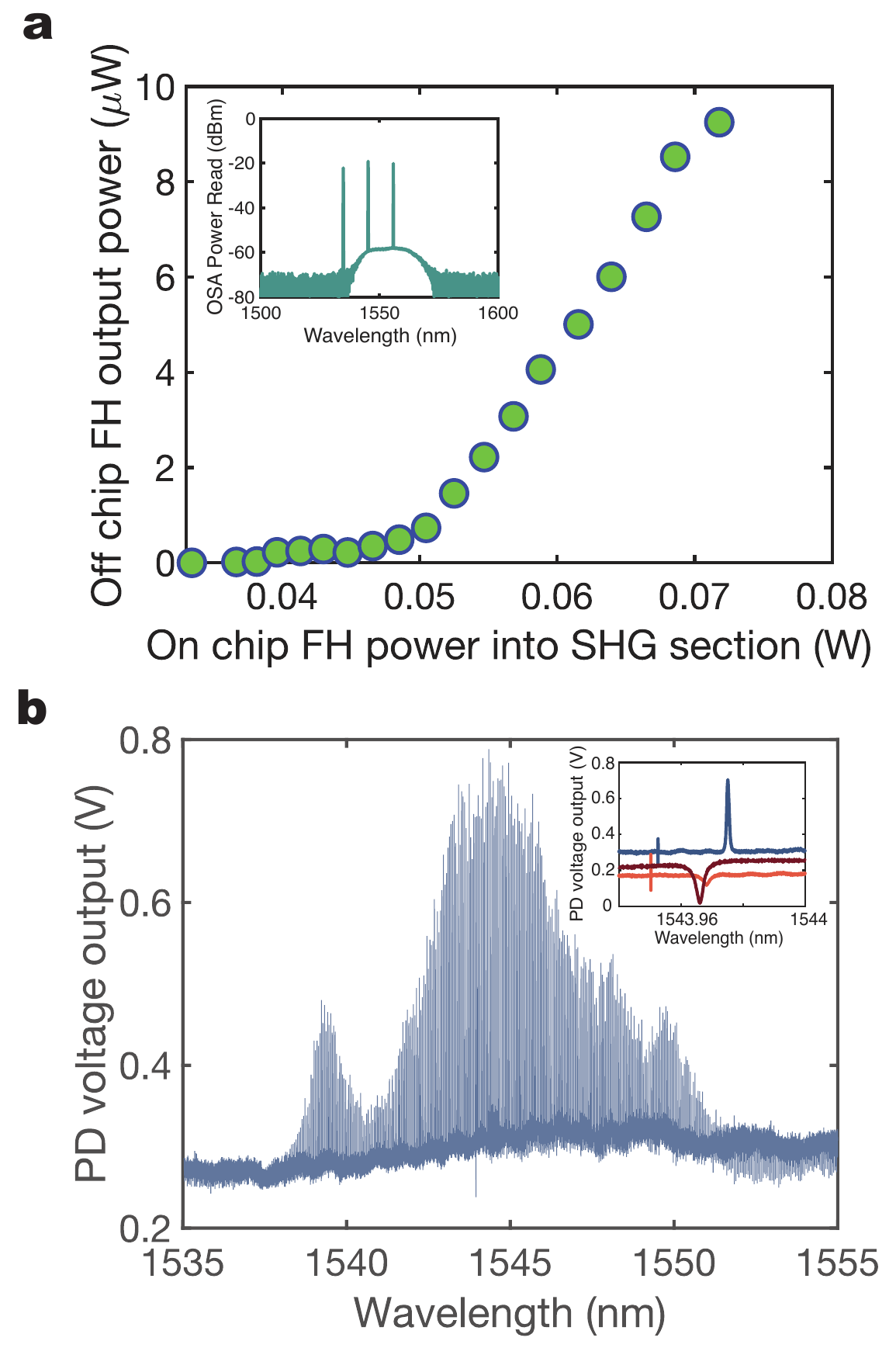}
  \end{center}
 \caption{
 \textbf{Optical parametric oscillator characterization. }
\textbf(a) Output fundamental harmonic (FH) power at the output lensed fiber vs. on-chip FH power going into the waveguide second harmonic generator (SHG) section. We begin to observe oscillation at approximately 50 mW going into the SHG section where the inset shows the optical spectrum analyzer (OSA, Yokogawa AQ6370D) trace of the oscillation. (b) Cavity transmission in the presence of parametric gain (see Supplementary Information). Inset shows transmission at different pump power levels around the wavelength of the FH light used to generate gain. Red is the cold cavity transmission, orange is the cavity transmission at gain/loss = 0.62, and blue is cavity transmission at gain/loss = 0.88. We estimate the gain/loss ratio based on the linewidth of the modes where they were 2.84 pm, 2.73 pm, and 0.9 pm for red, orange, and blue trace, respectively (see Supplementary Information). Sharp line on the left of the resonance corresponds to the beating of the two lasers.}
 \label{fig:fig2}
\end{figure}

A resonant cavity with parametric gain will oscillate once the gain exceeds the loss, forming an OPO. An OPO pumped such that its gain is lower than the oscillation threshold, will parametrically amplify incident fields and thus generate squeezing. In principle, the achievable squeezing in such a system is only limited by the escape efficiency -- the ratio between the rate at which photons that couple out of the device into the channel of interest divided by the total loss rate. Notably, while the amount of squeezing within the cavity is limited to at most 50\%, the emitted output from the cavity can display arbitrarily large amount of squeezing due to an interference effect \cite{yurke1984use, collett1984squeezing,gardiner1985input} in the limit of high escape efficiency. 

\begin{figure*}
  \begin{center} \includegraphics[width=\textwidth]{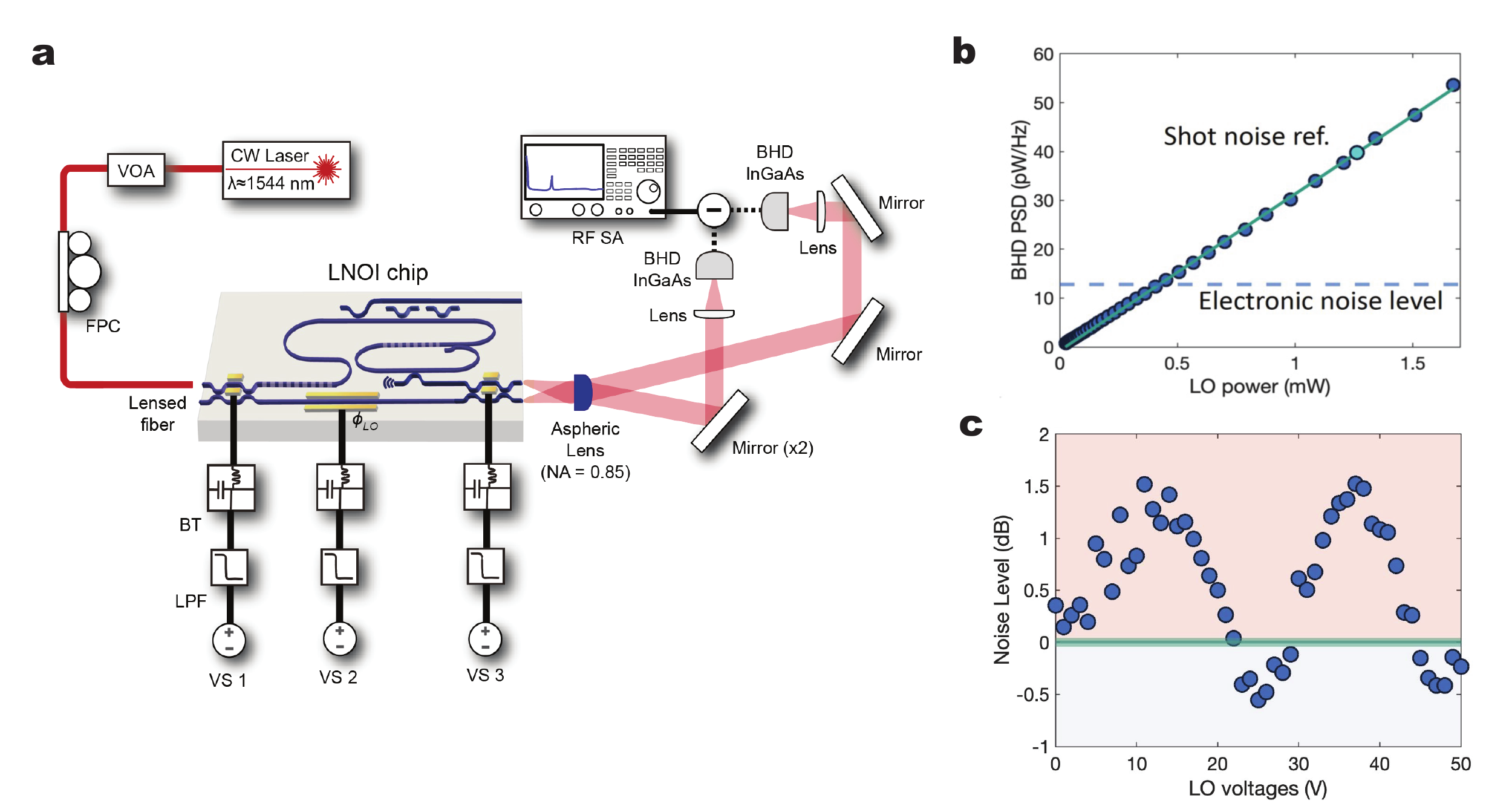}
  \end{center}
 \caption{
 \textbf{Squeezing measurement setup and results.}
\textbf{(a)} Squeezing measurement setup. Continuous wave light at 1544 nm from the low-noise laser is coupled into the chip using a lensed fiber. We control the input power going into the chip using a variable optical attenuator (VOA) and adjust the polarization using a fiber polarization controller (FPC). Three voltage sources are employed to control the phase of the light where we use low pass filter (LPF) and bias tee (BT) to suppress non-DC signals going into the electrodes. We use high numerical aperture (NA) aspheric lens to efficiently collect light from the two output waveguides. The two beams are routed and focused onto the InGaAs photodetectors of the balanced homodyne detector respectively. We measure the power spectral density (PSD) of the difference between the two photocurrents using the radio frequency spectrum analyzer (RFSA). \textbf{(b)} Measured PSD on the RFSA (electronic noise subtracted) averaged from 58 MHz to 60 MHz vs. LO power. \textbf{(c)} Measured PSD on the RFSA (electronic noise subtracted) averaged from 58 MHz to 60 MHz vs. DC voltage applied to the local oscillator (LO) phase shifter at the TEC temperature of 43.4 \textdegree C. On-chip fundamental harmonic power going into the SHG section was 20 mW.}
\label{fig:fig3}
\end{figure*}

Two critical properties of the OPO are its oscillation threshold and its parametric gain spectrum. We characterize both. First, by performing a power sweep of the fundamental harmonic light entering the SHG section, we estimate the oscillation threshold of our OPO. We send FH power onto  the chip, aligning its wavelength to the peak SHG response. This light is converted to SH, which then pumps the OPO. At sufficiently high power, the OPO oscillates and generates a significant amount of FH light. We measure the total power in this OPO-generated FH light and its spectrum while sweeping the input FH power. We observe significant output FH power above the oscillation threshold corresponding to an input FH power exceeding approximately 50 mW. This corresponds to an on-chip SH power of 25 mW inferred from an independent waveguide SHG nonlinear conversion efficiency measurement (see Supplementary Information). A typical OPO spectrum is shown in the inset of Fig 2a.

Secondly, we measure the gain spectrum of the OPO in the sub-threshold regenerative amplifier regime~\cite{siegman1986lasers, zhao2023large} by observing the amplification of a second incident laser tone that we sweep across the cavity modes. Under typical circumstances, with no gain, cavity resonances appear as dips in the transmission spectrum due to the incident optical power being lost through the intrinsic loss channels. However, increasing gain begins to counteract this intrinsic loss, altering the transmission spectrum in the process. When the gain entirely compensates for the \emph{intrinsic} loss within the cavity, the dips in the transmitted intensity vanish, signifying that the gain has compensated all loss. Any additional gain causes the amount of transmitted power to exceed the input, and thus inverts the cavity line shape, leading to a transmission \emph{peaks} with values surpassing unity. This is sometimes referred to as regenerative amplification~\cite{siegman1986lasers}. To make this measurement, we introduce a second laser through a separate grating coupler input path. This setup allows us to probe the transmission spectrum across different pump powers, while sweeping the wavelength of the second laser. We observe peaks indicative of amplification as shown in Fig. 2(b). The highest gain/loss ratio for modes in this plot is 0.88, as inferred from the mode's linewidth (see Supplementary Information). We show the effect of increasing gain on a single mode lineshape in the inset.


After characterizing the performance of the individual components, we carry out the squeezing measurement using the setup shown in Fig 3.(a). The output from the low-noise CW laser (ULN15PC, Thorlabs) goes through a variable optical attenuator (VOA) and a fiber polarization controller (FPC) before being coupled into the waveguide via a lensed fiber. We measure the coupling efficiency from the lensed fiber (SMF-28) to the input waveguide using a diagnostic waveguide (not shown in Fig 1.(b)), arriving at approximately $\sim$32\%, ignoring the propagation loss on-chip. Three voltage sources (PLH250-P, Aim-TTi) are used to apply DC voltages to the three electrodes on the chip respectively. We use external probes to connect the output of the voltage sources to the electrodes. We filter the voltages going to electrodes with low-pass filters (SLP-1.9+, Mini-Circuits) and bias tees (ZFBT-4R2GW+, Mini-Circuits). 

We collect the light emitted from the facets of the two output waveguides at the edge of the chip using a high numerical aperature (NA) aspheric lens (C037TME-D, Thorlabs). The two beams are collimated after the aspheric lens and propagate at slightly different angles. At a sufficient distance, the two beams are well separated. We use a dielectric mirror at this distance to route the two beams individually to each of the two InGaAs detectors of the BHD (HBPR-450M-10K-IN-FST, Femto). Finally we use a plano-convex lenses (LA1134-C, Thorlabs) before the detectors to focus down the beam to the size of the photodiode. The total transmission from the output waveguide to the InGaAs detector was estimated to be 70\%.

A precise shot noise reference is essential for calibrating the squeezing measurement. To obtain this reference, we collect the RF spectrum of the balanced homodyne detector under balanced conditions while varying the incident optical power. We use the setup identical to the squeezing measurement scheme (Fig 3.(a)) for the shot noise measurement. To perform this measurement, the light is coupled into the input waveguide and the TBS electrode is biased to send nearly all of the light into the LO path. The two optical powers incident on the two InGaAs detectors of the BHD detector are balanced using the BHD BS. A power sweep is carried out by varying the input FH power going into the chip.  In Fig 3. (b), the power sweep shows an ideal linear relation between the PSD and the incident optical power. Here, we average the power spectral density (PSD) measured on the radio frequency spectrum analyzer (RFSA) over 2 MHz window centered at around 59 MHz. For all of the squeezing measurements, we use the real-time mode on the RFSA (FSW 8, Rohde \& Schwarz) where we set the resolution bandwidth to 100 kHz and the video bandwidth to 28 MHz.

\begin{figure}
  \begin{center}
      \includegraphics[width=1\columnwidth]{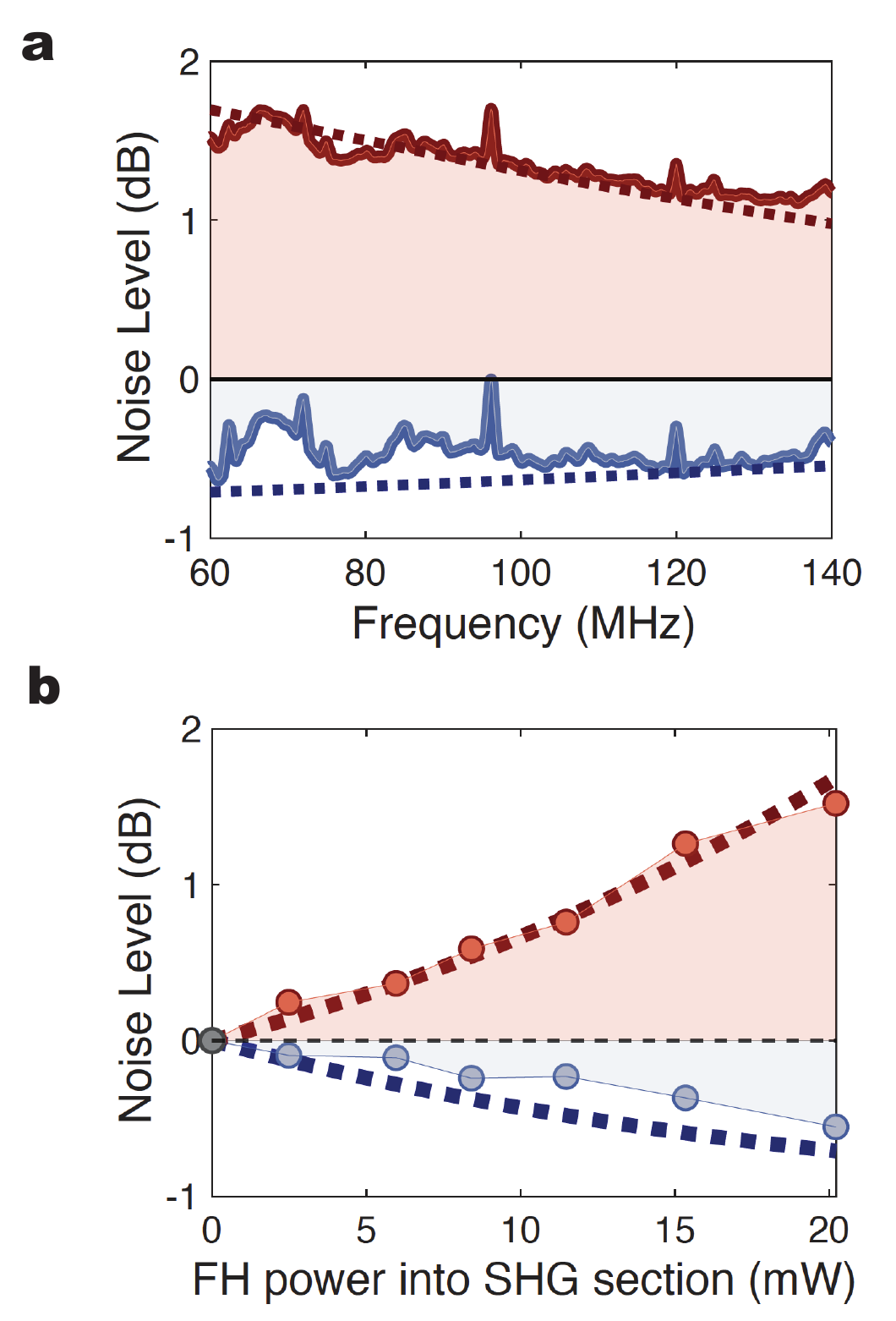}
  \end{center}
 \caption{
 \textbf{Squeezing spectra and power sweep. }
\textbf{(a)} Maximum (red) and minimum (blue) measured PSD spectrum (normalized by the shot noise, electronic noise subtracted) vs. measurement frequency of the RFSA at the TEC temperature setting of 43.4 \textdegree C with on-chip fundamental harmonic power of 20 mW going into the SHG section. Dotted lines represent the squeezing theory (see Supplementary Information). \textbf{(b)} Measured PSD values (normalized by shot noise, electronic noise subtracted) averaged from 58 MHz to 60 MHz at around the peak (red) and dip (blue) of the LO phase shifter DC voltage sweep vs. on-chip input FH power going into the SHG section. TEC temperature setting was at 43.4 \textdegree C for this measurement. Dotted lines are theory lines (see Supplementary Information).}
 \label{fig:fig4}
\end{figure}

Having calibrated the shot noise level, we generate squeezed light with the OPO and use the LO phase to vary the quadrature we measure to observe both squeezing and anti-squeezing. The LO power is fixed at 1.3 mW for this measurement. At this optical power, shot noise is 3.3 times larger than the total electronic noise level measured on the RFSA. The temperature setting of the thermoelectric cooler (TEC, Thorlabs TECF2S) is changed over 2.5 \textdegree C centered around 42.5 \textdegree C to go on and off the mode (see Fig. S2). From an independent cavity transmission measurement, we verify that tuning the TEC temperature setting by 1.2 \textdegree C tunes the mode spectrum over the free spectral range (FSR) of the cavity (Fig 1.(b)). Separately, we sweep the LO phase electro-optically by applying a voltage to the LO phase shifter. We step the TEC temperature and collect RF spectra at each LO phase. We observe squeezing and anti-squeezing when the mode resonance wavelength aligns with the laser wavelength (see Supplementary Information). At 20 mW of on-chip FH power going into the SHG section, we measure squeezing of 0.55 dB and anti-squeezing of 1.55 dB (Fig 3.(c)).

The bandwidth of squeezing and anti-squeezing from a cavity is limited by the linewidth of the mode of the cavity. In Fig 4.(a), as we show that we are detuned from the cavity resonance, the amount of squeezing and anti-squeezing decreases. At 20 mW of on-chip FH power going into the SHG section, we measure squeezing and anti-squeezing spectra from 60 MHz to 140 MHz. The frequency span of the measurement was limited by the RFSA. At frequencies below 60MHz, we observe excess noise and are no longer limited by shot noise. This measured spectra aligns with the squeezing theory (see Supplementary Information).

In Fig 4.(b), we perform a power sweep of the input FH and measure squeezing and anti-squeezing at different input powers. To perform this measurement, we first tune the TEC temperature to set the cavity frequency on resonance with the FH. At different input powers, we step the DC voltage applied to the LO phase shifter from 0 to 50V and collect the RF spectrum at each voltage. We plot the measured PSD values around the peak and dip of the swing averaged from the measurement frequency of 58 MHz to 60 MHz at different input powers. As we increase the input power, squeezing and anti-squeezing increase as expected (see Supplementary Information).


Our work demonstrates the generation and tomography of degenerate squeezed states using an integrated optical parametric oscillator on thin-film lithium niobate. We have created a monolithic photonic circuit that integrates almost all essential components, enabling 0.55 dB of measured squeezing with only 20 mW of optical pump power. This represents a substantial advance toward efficient and portable squeezed light sources by harnessing the versatile $\chi^{(2)}$ nonlinearities available in emerging integrated photonics. When combined with lasers and detectors, our device can be developed into a complete system for deployable quantum-enhanced optical sensors. Considerable improvements in squeezing levels are possible through further optimization of conversion efficiency, cavity $Q$, and escape efficiency and may enable the use of such devices in optical quantum computing \cite{madsen2022quantum,chen2014experimental,asavanant2019generation}. With rapid progress in this burgeoning field, integrated resonant squeezers like the one created here will help unlock the full potential of quantum light sources across diverse applications.

\section*{Data Availability}
The data sets generated during and/or analyzed during this study are available from the corresponding author on request.

\clearpage
\onecolumngrid
\section*{Supplementary Information}
\hfill \break
\twocolumngrid
\section*{Design of the quantum photonic circuit} 

The waveguide geometry of the photonic integrated chip used in the OPO squeezing experiment follows our previous work \cite{mckenna2022ultra}, with a ridge waveguide width of 1.2 um, height of 500 nm and etch of 300 nm. The length of the poled section for both SHG and OPA is 1 cm, and there is an additional diagnostic SHG waveguide in the same poled section that is 6.5 mm long (not shown in Fig 1.(b)). Dichroic beamsplitters, which are 400 $\mu$m in length, allow the FH light to evanescently couple into the 1-um-spaced adjacent waveguide while preventing the second harmonic light from traversing. The length of the racetrack resonator is approximately 2.2 cm, and it is designed to resonate the fundamental harmonic light and single-pass the second harmonic light (\textit{i.e.}, pump light). A dichroic beamsplitter closes the racetrack cavity for the FH and dumps the SH pump (Fig 1.(b)). The external coupler of the racetrack resonator is 120 $\mu$m in length and 1 $\mu$m in gap. The grating coupler is designed to operate at 1550 nm of maximum response at around the incident angle of 27 degrees. The beamsplitters (TBS and BHD BS) consist of couplers that are 250 $\mu$m in length, and the lengths of the arms are around 2.7 mm. We place 2.5 mm long electrodes next to one of the arms to induce phase shifts by applying DC voltages, and the gap between the waveguide and the electrode is 2.15 $\mu$m.

\section*{Fabrication of the quantum photonic circuit} 

We start with an X-cut LNOI (NanoLN) chip with 500-nm-thick LN to fabricate the quantum photonic circuit. We first deposit 100 nm of SiO\textsubscript{2} using a high density plasma enhanced chemical vapor deposition (Plasma Therm Versaline HDP CVD) tool. On top of the SiO2, we pattern Cr electrodes for periodic poling using electron beam lithography  (JEOL 6300-FS, 100-kV) based liftoff process. Applying high voltage pulses to these electrodes with a peak voltage of 700 V, we periodically pole the regions on the chip where SHG and OPA waveguides are placed. We were able to obtain consistent poling over the 1-cm-long section, verified by the SHG microscope (see Fig S3). After periodic poling, we remove the Cr electrodes using the Cr etchant and remove the 100-nm-thick oxide layer with buffered oxide etchant (BOE). We then pattern hydrogen silsesquioxane (HSQ) mask using electron beam lithography and etch the LN using an ion mill to make waveguides. We perform another electron beam lithography based liftoff process to pattern 100-nm gold electrodes next to the waveguides where we aim to control the phase of the transmitted light. Using the HDP CVD system, we deposit 700 nm of SiO\textsubscript{2} for cladding. Then we deposit photoresist (SPR3612) and make patterns using photolithography for vertical interconnect access (VIA). We etch the SiO\textsubscript{2} cladding down to the 100-nm gold electrodes to make vias using inductively coupled plasma etch system (Plasma Therm Versaline LL ICP Dielectric Etcher). After making vias, we pattern 200-nm-thick gold pads on the top surface of the chip to make contact with the probes. For this, we use the identical process that we used for making the 100-nm gold electrodes. Lastly, we anneal the chip in an oxygen environment at 500 \textdegree C for 8 hours.

\section*{Waveguide second harmonic generator characterization} 

To characterize the second harmonic generation section of our circuit, we start by measuring its spectral response. To access the waveguide SHG section, we couple light into the output port of the first DBS and collect the output at the top input waveguide (Fig 1.(b)). Since the coupler of the TBS does not couple SH light to an adjacent waveguide, we may take a wavelength scan of the fundamental harmonic to look at the spectral response at the second harmonic. In Fig S3.(a), we show the spectral response of the waveguide SHG section measured with this method. To extract the nonlinear conversion efficiency, we use the additional diagnostic SHG waveguide (not shown in Fig 1.(b)). We perform calibrated power sweep at the wavelength that gives the peak SHG response. We plot the SH power on-chip versus FH power on chip in Fig S3.(b) where the quadratic fit gives normalized conversion efficiency of 1000\%/W. Assuming negligible measurable propagation loss of the diagnostic waveguide, the SMF-28 lensed fiber to waveguide coupling efficiency was 32\% at 1550 nm. Waveguide to SMF-28 lensed fiber coupling efficiency was 30\% at 780 nm. Although the peak SHG response is near 1543 nm, we must consider the nonlinear conversion efficiency at the wavelength of our low-noise laser used for the squeezing measurement. Fig S3.(c) shows the normalized conversion efficiency near the wavelength of the low-noise laser (1544.4 nm). Assuming that the nonlinear conversion efficiency at the peak is identical for both the diagnostic SHG waveguide and the actual SHG waveguide, we can infer the normalized conversion efficiency based on the relative spectral response at different wavelengths. The normalized conversion efficiency was around 600\%/W at 1544.4 nm using this analysis. We also observe that the SHG spectral response does not shift significantly as we change the TEC temperature setting over 2 \textdegree C, which is the temperature tuning range of our squeezing measurement.

\section*{Cavity mode characterization} 

To measure the cavity transmission, we couple light from the laser at the fundamental harmonic into the grating coupler using an angle-cleaved fiber (Fig S5.(a)). We scan the wavelength of the tunable laser (Santec, TSL710) from 1500 nm to 1620 nm and collect the transmission. The transmission of the cavity is shown in Fig S5.(a), where the broad spectral response is from the grating coupler. Assuming that the cavity is undercoupled, a lorentzian fit to the mode near the low-noise laser wavelength gives $Q_\text{int}$ of 950k and $Q_\text{tot}$ of 550k. To determine the coupling condition of the cavity mode at the operating wavelength, we measure the phase response across the optical resonance~\cite{chan2012laser,herrmann2022mirror}. We park the laser near the optical resonance frequency and generate optical sidebands using an electro-optic modulator. Then, we measure the phase response of the optical to optical scattering parameter with the vector network analyzer (VNA). Shown in Fig S4 (b), the measured phase response shows relatively small change in phase across the resonance frequency, indicating that the mode is undercoupled. By fitting the measured phase response data, we verify that the mode is undercoupled.

\section*{Optical parametric oscillator characterization} 

We characterize the optical parametric oscillator using the measurement setup shown in Fig S5.(a). First, we estimate the threshold of our OPO using tunable laser 1. We vary the FH power that couples into the input waveguide and take a broad wavelength scan in the range of the EDFA gain (1530 nm to 1560 nm) at different input on-chip power levels. For each scan, we collect the voltage read on the InGaAs avalanche photodiode (APD, Thorlabs APD410C). We observe an abrupt increase in the voltage read on the APD when the OPO starts to oscillate. At the wavelength exhibiting the minimum oscillation threshold, we perform a high-resolution wavelength sweep to record the optical spectrum (inset of Fig. 2a). Next, we measure the cavity transmission in the presence of parametric gain using two CW lasers at the fundamental harmonic. To generate parametric gain in the racetrack resonator, we park the wavelength of tunable laser 1 at the peak waveguide SHG response. We tune the magnitude of parametric gain by changing the input fundamental harmonic power going into the chip with the VOA. At different levels of parametric gain, we take a wavelength scan using tunable laser 2 and collect the voltage read on the InGaAs APD. For the cavity transmission measurement, we use an angle-cleaved fiber to couple telecom light from the tunable laser 2 into the grating coupler. Fig S5.(b) shows the theory plot of the cavity transmission in the presence of parametric gain. As we increase the gain, the linewidth of the mode becomes narrower and at a high gain region below threshold we observe peak due to amplification.

\section*{Electro-optic phase shifter characterization} 

To characterize the electro-optic phase shifter, we take broad wavelength scans at different DC voltages applied to the TBS electrode. We couple light from a tunable laser at the fundamental harmonic into the bottom input waveguide using a lensed fiber. Then, we step the voltage from 0 to 50V and collect the output light using a lensed fiber from the BHD BS top output waveguide and the BHD BS bottom output waveguide in separate measurements. Assuming identical coupling efficiencies for both output waveguides, we add the two output powers to obtain the magnitude at the LO path after the TBS. In Figure 1. (b), we plot this sum at different voltages at a wavelength of 1520 nm and normalize by the peak transmission value to obtain $V_{\pi}$.

\section*{Fundamental harmonic transmission characterization of the dichroic beamsplitter} 

The dichroic beamsplitter (DBS) consists of a directional coupler with identical waveguides adjacent to each other. The length of this coupler is 400 $\mu$m and it is designed to transfer light at the telecom wavelength with 100\% efficiency. The gap is set to 1 $\mu$m to prevent second harmonic light from evanescently coupling to the adjacent waveguide at the coupler section. The coupling coefficient of the directional coupler is wavelength-dependent, and the performance of the DBS deviates from 100\% transmission at the low-noise laser wavelength (1544.4 nm) due to the fabrication imperfection. To characterize the transmission of the DBS at the fundamental harmonic, we couple light using a SMF-28 lensed fiber into the output waveguide from the first DBS (see Fig 1.(b)). Then, we use a tunable laser (TSL-710, Santec) and take a broad wavelength scan. We collect light coming out from the two input waveguides, respectively, using an SMF-28 lensed fiber. To get the power at the waveguide between the TBS and the DBS, we add the optical power at the two input waveguides. For this, we assume that the coupling efficiencies from both input waveguides to a lensed fiber are identical. In Fig S6.(a), the wavelength dependence of the inferred optical power at the waveguide after the light passes through the dichroic beamsplitter is shown. Here, the wavelength dependence of the coupling efficiency (measured from an independent measurement using the diagnostic SHG waveguide) is taken into account when we infer the power at the waveguide. We estimate the transmission of the DBS to be 94$\pm$2\% at the low-noise laser wavelength, assuming 100\% transmission at around 1580 nm. In Fig S6.(b), we show that the wavelength dependence of the coupling efficiency is minimal and that the wavelength dependence in Fig S6.(a) is due to DBS. 

\section*{Laser noise characterization} 

We study the noise properties of the low-noise laser prior to our squeezing measurement. First, we characterize the intensity noise of the laser. To check that our laser does not have excess intensity noise, we take a power sweep and measure the photocurrent PSD from the photodetector at different incident optical powers using RFSA (see S.I.). If there is no excess intensity noise, the measured power spectral density (PSD) should have a linear dependence with respect to the incident optical power. Fig S7.(d) shows this linear relation, verifying that the laser has no measurable excess intensity noise. The measured photocurrent PSD spectrum at different incident optical power on the photodetector is shown in Fig S7.(c). Next, we characterize the phase noise of the laser. To measure the phase noise of the laser, we send the output of the laser to one of the input port of a 2 by 2 fiber Mach-Zehnder interferometer (MZI) with a known FSR of 67 MHz. One of the two output ports is routed to the photodetector and we measure the photocurrent PSD spectrum using the RFSA. Shown in Fig S7.(e), the measured phase noise aligns with the theory that assumes a laser linewidth of 100 Hz, which is the reported value from the manufacturer.

\section*{Loss in our measurement and the anticipated on-chip squeezing with improved components} 

The measured squeezing in our experiment is limited by the loss in our OPO resonator, the loss after the output waveguide, and the non-ideal nonlinear conversion efficiencies on the chip. Our racetrack resonator has intrinsic quality factor of 950k where intrinsic quality factors of 10 million have been demonstrated on the same platform \cite{zhang2017monolithic}. There are two major sources of loss after the output waveguide: insertion loss of the aspheric lens and low quantum efficiencies of the photodiodes of the balanced homodyne detector. The insertion loss of our aspheric lens is over 20\% at the telecom wavelength due to the material absorption (BD-2 glass). Using an aspheric lens with a material that has lower loss at the telecom wavelength (e.g., N-F2) and smaller f-number, we expect the coupling efficiency to be over 90\%. The quantum efficiencies of the InGaAs photodiodes of our balanced homodyne detector were around 75\% at the telecom wavelength, where balanced homodyne detectors with InGaAs photodiodes of quantum efficiencies over 99\% have been utilized in squeezing experiments \cite{vahlbruch2016detection}. Normalized conversion efficiency at the peak SH response was around 1000\%/W for a 1-cm-long PPLN waveguide on our chip. The spectral response illustrated in Fig S3 (a) exhibits a wider distribution and diminished peak efficiency compared to theory. By analyzing the integral of this response, we infer that minimizing inhomogeneity could potentially improve the peak efficiency to 3000\%/W-cm$^2$, where efficiency of 4000\%/W-cm$^2$ is theoretically feasible \cite{wang2018ultrahigh}. Fig S8 shows anticipated on-chip squeezing and anti-squeezing assuming achievable performance improvements of each of the components. Assuming normalized conversion efficiency of 4000\%/W-cm$^2$, \textit{Q}\textsubscript{int} of 10 million and \textit{Q}\textsubscript{tot} of 200$k$, we expect on-chip squeezing of 16 dB and on-chip anti-squeezing of 23 dB at the on-chip input fundamental harmonic power of 26 mW going into the waveguide SHG section from an OPO with identical geometry used in this experiment.

\section*{Squeezing theory} 

\subsection*{System hamiltonian and equations of motion}
The hamiltonian for a periodically poled doubly resonant cavity that consists of 2\textit{N}+1 modes around the fundamental frequency ($\omega_p$) with a non-resonant pump at the second harmonic could be written as,
\begin{eqnarray}
    \frac{\hat{H}}{\hbar}=&&2\omega_p\hat{b}^{\dagger}\hat{b} + \sum_{n=-N}^{n=N}\omega_n\hat{a}^{\dagger}_n \hat{a}_n + \sum_{n=0}^{n=N}(g_n\beta_n e^{-2i\omega_p t}\hat{a}^{\dagger}_n\hat{a}^{\dagger}_{-n} \nonumber\\
&& +g_n\beta_n e^{2i\omega_p t}\hat{a}_n\hat{a}_{-n})
\label{eq:eqM1}
\end{eqnarray}

\noindent where $\omega_0$ is the cavity resonance frequency closest to the fundamental frequency ($\omega_p$). $g$ is the nonlinear interaction rate and $\beta$ is the pump field (for $n=0$, $\beta_n \rightarrow 2\beta_0$). The Heisenberg equation of motion including the loss terms is then,
\begin{eqnarray}
    \frac{d\hat{a}_n}{dt}=&&-i\omega_n\hat{a}_n-ig_n\beta_n e^{-2i\omega_pt}\hat{a}^{\dagger}_{-n}-\frac{\kappa_n}{2}\hat{a}_n \nonumber\\
&& +\sqrt{\kappa_{n,e}}\hat{a}_{in}(t)  +\sqrt{\kappa_{n,i}}\hat{a}_{in,i}(t).
\label{eq:eqM2}
\end{eqnarray}

\noindent We now go into the frame of the fundamental harmonic where $\hat{a}_n \rightarrow \hat{a}_ne^{-i\omega_pt}$ and $\hat{a}^{\dagger}_n \rightarrow \hat{a}^{\dagger}_ne^{i\omega_pt}$. Equations of motion then becomes
\begin{eqnarray}
    \frac{d\hat{a}_n}{dt}=&&-i(\omega_n-\omega_p)\hat{a}_n-ig_n\beta_n \hat{a}^{\dagger}_{-n}-\frac{\kappa_n}{2}\hat{a}_n \nonumber\\
&& +\sqrt{\kappa_{n,e}}\hat{a}_{in}(t)e^{i\omega_pt}+\sqrt{\kappa_{n,i}}\hat{a}_{in,i}(t)e^{i\omega_pt}.
\label{eq:eqM3}
\end{eqnarray}

\noindent Our cavity has equally spaced modes in the frequency domain (assuming negligible GVD over the frequency range of interest). To capture this mode structure in our model, we introduce $\Omega_n = n\Omega$, where $\Omega$ is the free spectral range (FSR) of the cavity. Defining $\Delta_n$ as the amount of detuning of the $n^\text{th}$ mode from the resonance frequency, we have $\omega_n = \Omega_n + \Delta_n + \omega_p$. To understand the dynamics for each of the frequency modes, we think in terms of the slowly varying operators which we define as $\hat{a}_{n,s}(t) = e^{i\Omega_n t}\hat{a}_n(t)$ and $\hat{a}_{in,s}(t) = \hat{a}_{in}(t)e^{i\omega_p t}$. Then, for slowly varying operators, the equation of motion becomes
\begin{eqnarray}
    \frac{d\hat{a}_{n,s}}{dt}=&&-i\Delta_n\hat{a}_{n,s} -ig_n\beta_n\hat{a}^{\dagger}_{-n,s}-\frac{\kappa_n}{2}\hat{a}_{n,s} \nonumber\\&& +\sqrt{\kappa_{n,e}}\hat{a}_{in,s}(t)e^{in\Omega t} +\sqrt{\kappa_{n,i}}\hat{a}_{in,i,s}(t)e^{in\Omega t} \nonumber\\=&&-i\Delta_n\hat{a}_{n,s} -ig_n\beta_n\hat{a}^{\dagger}_{-n,s}-\frac{\kappa_n}{2}\hat{a}_{n,s} \nonumber\\&&+\sqrt{\kappa_{n,e}}\hat{a}_{in,s,n}+\sqrt{\kappa_{n,i}}\hat{a}_{in,i,s,n}.
\label{eq:eqM4}
\end{eqnarray}

In our squeezing measurement, we are interested in the mode closest to the frequency of the local oscillator ($\omega_p$). Since the bandwidth of our homodyne detector (450 MHz) is much smaller than the FSR of our cavity ($\sim$5.7 GHz), in the frame of local oscillator frequency the detector cannot capture the dynamics of other frequency modes because it is rapidly oscillating. 

Now, we consider the equation of motion for the single mode closest to the frequency of $\omega_p$. Dropping the subscript for the simplicity of notation we get, 
\begin{eqnarray}
    \frac{d\hat{a}}{dt}=&&-i\Delta\hat{a}-2ig\beta \hat{a}^{\dagger}-\frac{\kappa}{2}\hat{a} +\sqrt{\kappa_{e}}\hat{a}_{in}+\sqrt{\kappa_{i}}\hat{a}_{in,i}.
\label{eq:eqM5}
\end{eqnarray}
In the frequency domain, the equation of motion becomes
\begin{eqnarray}
    -i\omega\hat{a}(\omega) = -i\Delta\hat{a}(\omega)-2ig\beta\hat{a}^{\dagger}(-\omega) && \nonumber\\ -\frac{\kappa}{2}\hat{a}(\omega)+\sqrt{\kappa_{e}}\hat{a}_{in}(\omega)+\sqrt{\kappa_{i}}\hat{a}_{in,i}(\omega)
\label{eq:eqM6}
\end{eqnarray}
\begin{eqnarray}
    i\omega\hat{a}^{\dagger}(\omega) = i\Delta\hat{a}^{\dagger}(\omega)+2ig\beta^*\hat{a}(-\omega) && \nonumber\\ -\frac{\kappa}{2}\hat{a}^{\dagger}(\omega)+\sqrt{\kappa_{e}}\hat{a}_{in}^{\dagger}(\omega)+\sqrt{\kappa_{i}}\hat{a}_{in,i}^{\dagger}(\omega)
\label{eq:eqM7}
\end{eqnarray}
where Eq.(7) is the hermitian conjugate form of Eq.(6).

\subsection*{Squeezing at the output of the cavity} 
We follow the derivations from \cite{collett1984squeezing, gardiner1985input} and calculate squeezing in the good cavity limit (i.e., $\kappa \sim \kappa_e$). We will include effect of intrinsic loss ($\kappa_i$) in the escape efficiency factor defined in the following section. Using the boundary condition $\hat{a}_{out}(\omega) = \sqrt{\kappa}\hat{a}(\omega)-\hat{a}_{in}(\omega)$ and the equations of motion (Eq. (6) and Eq. (7)), we could express $\hat{a}_{out}(\omega)$ in terms of $\hat{a}_{in}(\omega)$
\begin{multline}
    \hat{a}_{out}(\omega) = \\ \frac{\frac{\kappa^2}{4}+\omega^2+4g\lvert \beta \rvert^2}{(\frac{\kappa}{2}-i\omega)^2-4g^2\lvert \beta \rvert ^2}\hat{a}_{in}(\omega)- \frac{2ig\beta\kappa}{(\frac{\kappa}{2}+i\omega)^2-4g^2\lvert \beta \rvert ^2}\hat{a}_{in}^{\dagger}(-\omega)
\label{eq:eqM8}
\end{multline}

\begin{multline}
    \hat{a}_{out}^{\dagger}(\omega) = \\ \frac{\frac{\kappa^2}{4}+\omega^2+4g\lvert \beta \rvert^2}{(\frac{\kappa}{2}+i\omega)^2-4g^2\lvert \beta \rvert ^2}\hat{a}_{in}^{\dagger}(\omega) + \frac{2ig\beta^*\kappa}{(\frac{\kappa}{2}+i\omega)^2-4g^2\lvert \beta \rvert ^2}\hat{a}_{in}(-\omega).
\label{eq:eqM9}
\end{multline}

\noindent We define the two quadratures of the output as $\hat{X}_{out}=\hat{a}_{out}^{\dagger}+\hat{a}_{out}$ and $\hat{Y}_{out}=i(\hat{a}_{out}^{\dagger}-\hat{a}_{out})$. The variance of $\hat{X}_{out}$ is  $\langle \hat{X}_{out}(\omega), \hat{X}_{out}(\omega') \rangle = \langle \hat{X}_{out}(\omega) \hat{X}_{out}(\omega') \rangle  - \langle \hat{X}_{out}(\omega)\rangle \langle\hat{X}_{out}(\omega')\rangle$. Using $\langle \hat{X}_{out}(\omega)\rangle = 0$ and defining the phase of the output as $\phi_{out}$ (i.e., $\hat{a}_{out}(\omega)\rightarrow e^{i\phi_{out}}\hat{a}_{out}(\omega)$) we get

\begin{eqnarray}
 \langle \hat{X}_{out}(\omega), \hat{X}_{out}(\omega') \rangle = && \langle \hat{X}_{out}(\omega) \hat{X}_{out}(\omega') \rangle \nonumber \\  = && e^{-2i\phi_{out}}\langle \hat{a}_{out}^{\dagger}(\omega)\hat{a}_{out}^{\dagger}(\omega')\rangle \nonumber \\ && +
 e^{2i\phi_{out}}\langle \hat{a}_{out}(\omega)\hat{a}_{out}(\omega')\rangle \nonumber \\ && +
 2\langle \hat{a}_{out}^{\dagger}(\omega)\hat{a}_{out}(\omega')\rangle \nonumber \\ && +
 \delta(\omega-\omega')
\label{eq:eqM10}
\end{eqnarray}
where we used $[\hat{a}_{out}(\omega), \hat{a}_{out}^{\dagger}(\omega')] = \delta(\omega-\omega')$. We can calculate each of the terms in Eq.(10) and get
\begin{multline}
    \langle \hat{a}_{out}(\omega)\hat{a}_{out}(\omega') \rangle = \\ \frac{-2ig\beta\kappa\big[\frac{\kappa^2}{4}+\omega^2+4g\lvert \beta \rvert^2\big]}{\big[(\frac{\kappa}{2}-i\omega)^2-4g^2\lvert \beta \rvert ^2\big]\big[(\frac{\kappa}{2}-i\omega')^2-4g^2\lvert \beta \rvert^2\big]}\delta(\omega+\omega')
\label{eq:eqM11}
\end{multline}

\begin{multline}
    \langle \hat{a}_{out}^{\dagger}(\omega)\hat{a}_{out}^{\dagger}(\omega') \rangle = \\ \frac{2ig\beta^*\kappa\big[\frac{\kappa^2}{4}+\omega'^2+4g\lvert \beta \rvert^2\big]}{\big[(\frac{\kappa}{2}+i\omega)^2-4g^2\lvert \beta \rvert ^2\big]\big[(\frac{\kappa}{2}+i\omega')^2-4g^2\lvert \beta \rvert^2\big]}\delta(\omega+\omega')
\label{eq:eqM12}
\end{multline}

\begin{multline}
    \langle \hat{a}_{out}^{\dagger}(\omega)\hat{a}_{out}(\omega') \rangle = \\ \frac{4g^2\lvert \beta \rvert^2\kappa^2}{\big[(\frac{\kappa}{2}+i\omega)^2-4g^2\lvert \beta \rvert ^2\big]\big[(\frac{\kappa}{2}-i\omega')^2-4g^2\lvert \beta \rvert^2\big]}\delta(\omega-\omega').
\label{eq:eqM13}
\end{multline}

Plugging in Eq.(11-13) to Eq.(10) and integrating over $\omega'$, we arrive at

\begin{multline}
S_{XX}(\omega) = \\ 1 +\frac{g\lvert \beta \rvert \kappa}{(\frac{\kappa}{2}-2g\lvert \beta \rvert)^2+\omega^2}(2+2\text{sin}(2\phi_{out}+\phi_{\beta})) \\ 
+ \frac{g\lvert \beta \rvert \kappa}{(\frac{\kappa}{2}+2g\lvert \beta \rvert)^2+\omega^2}(-2+2\text{sin}(2\phi_{out}+\phi_{\beta}))
\label{eq:eqM14}
\end{multline}

\noindent where we defined $\beta = \lvert \beta \rvert e^{i\phi_{\beta}}$. We can see that we could get minimum and maximum at $2\phi_{out}+\phi_{\beta}=\pm\frac{\pi}{6}$. Since the threshold condition for oscillation implies $\lvert \beta \rvert_{thr} = \frac{\kappa}{4g}$, we can further simplify our form into

\begin{equation}
S_{\pm}(\omega) = 1 \pm \frac{4\frac{\lvert \beta \rvert}{\lvert \beta \rvert_{thr}}}{(1\mp\frac{\lvert \beta \rvert}{\lvert \beta \rvert_{thr}})^2+4(\frac{\omega}{\kappa})^2}.
\label{eq:eqM15}
\end{equation}

\noindent Here, $+$ and $-$ indicates anti-squeezing and squeezing respectively. 

\subsection*{Including the loss in our system and its effect on measured squeezing} 

Loss after the squeezing section diminishes the amount of squeezing and anti-squeezing that could be measured. For every component in our system with optical transmission less than unity ($\eta<1$), we define an equivalent optical beamsplitter with the splitting ratio $\eta:(1-\eta)$ where the output is $\eta$ times the input and ($1-\eta$) times vacuum. The detected field quadrature after this beamsplitter could then be written as $\hat{X}_{out,det} = \sqrt{\eta}\hat{X}_{out}+\sqrt{1-\eta}\hat{X}_{vac}$. The source of vacuum noise is uncorrelated with the cavity output, and thus the field quadrature spectral density is $S_{XX,det}(\omega) = \eta S_{XX, out}(\omega)+(1-\eta)S_{XX,vac}(\omega)$ where $S_{XX,vac}(\omega) = 1$ is the shot noise. 

Our system has three major sources of loss: inefficient extraction efficiency of the cavity (low escape efficiency), transmission at the output of the chip to free space interface, and quantum efficiency of the photodetector. We define an equivalent beamsplitter for each of these components where the transmission coefficient are $\rho = \frac{\kappa_{e}}{\kappa}$ (escape efficiency), $T$ (transmission from the output of the cavity to the detector), and $\epsilon$ (quantum efficiency of our detectors). Then, we may write $S_{\pm, meas}(\omega) = \eta_{tot}S_{\pm}(\omega)+(1-\eta_{tot})$ where $\eta_{tot} = \rho T \epsilon$ following from the product of beamsplitters. Finally, from Eq. (15) we get the form for the measured squeezing and antisqueezing 

\begin{eqnarray}
    S_{\pm, meas} =&& 1\pm \rho T \epsilon \frac{4\sqrt{\frac{P}{P_{th}}}}{(1\mp\sqrt{\frac{P}{P_{th}}})^2+4(\frac{\omega}{\kappa})^2}.
\label{eq:eqM16}
\end{eqnarray}

\section*{Laser noise theory and measurement} 
\subsection*{Theory of laser noise}
The classical field amplitude of the output from a laser can be written in the general form,

\begin{eqnarray}
    \alpha(t) = \alpha e^{-i\omega_Lt}(1+N(t))e^{i\phi(t)} 
\label{eq:eqM17}
\end{eqnarray}

\noindent where $\omega_L$ is the frequency of the laser, $N(t)$ and $\phi(t)$ are related to the intensity noise and phase noise respectively.

For a laser without excess intensity noise (i.e., $N(t)=0$), we could relate the power spectral density of the phase noise term $\phi(t)$ to the linewidth of the laser. To see this, we calculate the autocorrelation of the classical field amplitude. Assuming that the random process $\phi(t)$ follows a gaussian probability distribution \cite{lax1967classical}, we can write the autocorrelation as,

\begin{align}
    \langle \alpha(t)\alpha^*(0) \rangle &= \lvert \alpha \rvert^2 \langle e^{i(\phi(t)-\phi(0)}\rangle \nonumber \\
    &= \lvert \alpha \rvert^2  e^{-\langle(\phi(t)-\phi(0)\rangle^2/2}
\label{eq:eqM18}
\end{align}

\noindent where we used the property of the moments of gaussian probability distribution. We can further simplify the variance in terms of the power spectral density of the phase fluctuations,

\begin{align}
    \langle(\phi(t)-\phi(0))^2\rangle &= \langle(\phi(t)^2+\phi(0)^2-2\phi(t)\phi(0)\rangle \nonumber \\
    &= 2\langle \phi(0)^2 \rangle -2\langle \phi(t)\phi(0)\rangle \nonumber \\
    &= \frac{1}{\pi}\int^{\infty}_{-\infty}S_{\phi\phi}(\omega)(1-e^{-i\omega t})d\omega \nonumber \\
    &= \frac{2}{\pi}\int^{\infty}_{-\infty}S_{\phi\phi}(\omega)\text{sin}^2(\omega t/2)d\omega.
\label{eq:eqM19}
\end{align}

We may consider the case where the frequency noise is white. Defining the power spectral density of the frequency noise as $S_{\delta\delta}(\omega) = C$ where $C$ is constant, the variance of the phase drift becomes
\begin{align}
\langle(\phi(t)-\phi(0))^2\rangle &= C\lvert t \rvert.
\label{eq:eqM20}
\end{align}
 We note that the frequency noise and the phase noise have the relation $\omega^2 S_{\delta\delta}(\omega) = S_{\phi\phi}(\omega)$.

\noindent The power spectrum is then
\begin{align}
S_{\alpha\alpha}(\omega) &= \int_{-\infty}^{\infty} \langle{\alpha(\tau)\alpha^\ast(0)}\rangle e^{i\omega\tau}d\tau \\
&=\int_{-\infty}^{\infty} |\alpha|^2 e^{-\frac{C}{2}|t|}e^{i\omega\tau}d\tau \\
&= |\alpha|^2 \frac{4C}{C^2+4\omega^2}
\label{eq:eqM21}
\end{align}

\noindent where the full-width at half-maximum (FWHM) is $C$, and the linewidth of the laser is $C/2\pi$ (Hz).

\subsection*{Intensity noise measurement}
To check if there is no measurable excess intensity noise in addition to the shot noise from an ideal laser, we may perform a power sweep of the output of the laser and measure the power spectral density (PSD) of the photodiode at different incident optical powers using the radio-frequency spectrum analyzer (RFSA) (Fig S7.(a)). The photocurrent operator of the photodiode when $\alpha(t)$ is incident with quantum fluctuation $\hat{a}_{vac}(t)$ is
\begin{align}
\hat{I}(t) &= G(\hat{a}_{vac}(t)+\alpha(t))^{\dagger}(\hat{a}_{vac}(t)+\alpha(t))
\label{eq:eqM22}
\end{align}
where $G$ is the gain of the detector ($G$ is equal to $e$ for an ideal photodetector with no amplification). Setting $G=1$ for simplicity, the photocurrent power spectral density can then be written as

\begin{align}
S_{II}(\Omega) &= \int_{-\infty}^{\infty}e^{i\omega \tau}\big<\hat{I}^{\dagger}(t)\hat{I}(t-\tau)\big>d\tau \nonumber \\ &= \eta^2|\alpha|^2+ \eta^2\int^\infty_{-\infty}e^{i\omega\tau}|\alpha(t)|^2|\alpha(t-\tau)|^2  d\tau \\ &= \eta^2|\alpha|^2+ \eta^2|\alpha|^4 S_{NN}(\Omega)
\label{eq:eqM23}
\end{align}

\noindent where $S_{NN}(\Omega)$ is the PSD of the intensity noise fluctuations and $\eta$ represents the non-unity efficiency of the detector. From Equation (26), we could see that the excess intensity noise has a quadratic dependence with respect to optical power incident on the photodetector. If there is measureable excess intensity noise on our laser, measured PSD of the photocurrent versus optical power incident on the detector would deviate from the ideal linear relation. In Fig S7.(d), we show that the power sweep shows a linear relation with coefficient of determination of $R^2=1$, verifying that our laser has no measureable excess intensity noise. 

\subsection*{Phase noise measurement}

To obtain an estimate of the phase noise of our laser, we send the output of our laser to a 2$\times$2 fiber Mach-Zehnder interferometer (MZI) (Fig S7. (b)). The fiber MZI introduces a delay to the light passing through one of the arms due to the path length mismatch between the two arms. In the case where phase noise is small ($e^{i\phi(t)}\sim 1+i\phi(t)$), we may write the output from one of the two output ports as

\begin{align}
    \alpha_{out}(t) = \alpha\times (1+i\phi(t)+e^{i\omega_LL/c}(1+i\phi(t+L/c))
\end{align}

\noindent where $L$ is the path length difference and $c$ is the speed of light. From equation (25), the power spectral density of the photocurrent of the detector can be written as
\begin{multline}
   S_{II}(\Omega) = \eta^2|\alpha_{out}|^2 \\ +\eta^2\int^\infty_{-\infty} |\alpha_{out}(\Omega)|^2|\alpha_{out}(\Omega')|^2 d\Omega'
\end{multline}

\noindent where we used the $S_{AA}(\Omega) = \int_{-\infty}^{\infty}e^{i\omega \tau}A^{*}(t)A(t-\tau)d\tau = \int_{-\infty}^{\infty}A^{*}(\Omega)A(\Omega')d\Omega'$. To get $|\alpha_{out}(\Omega)|^2$, we first calculate $|\alpha_{out}(t)|^2$ and then take the fourier transform. Assuming $\phi^2\ll \phi$, we arrive at the photocurrent power spectral density 
\begin{multline}
S_{II}(\Omega) = \eta^2|\alpha_{out}|^2+ \\ \eta^2|\alpha|^4 16\sin^2(\omega_LL/c)\sin^2(\Omega L/2c)S_{\phi\phi}(\Omega)
\end{multline}
where $S_{\phi\phi}(\Omega)$ is the PSD of the phase fluctuation. The shot noise level for this measurement is $\eta^2|\alpha_{out}|^2$ and it has a fixed relation set by the transfer function of the fiber MZI, $|\alpha_{out}|^2 = |\alpha|^2\sin^2(\omega_LL/2c)$. Since we know the free spectral range (FSR) of our fiber MZI (67 MHz) from an independent measurement, by normalizing the PSD by the shot noise we can extract $S_{\phi\phi}(\Omega)$.

In a laboratory environment, fiber length may fluctuate and thus getting an accurate shot noise reference could be important for the phase noise measurement described in this section. We collect the DC voltage readout on the high-speed photodetector (Newport 12 GHz) when we take the PSD and use this as the reference for the shot noise. Figure S7.(e) compares the measured phase noise PSD to the theoretical prediction using the manufacturer-specified 100 Hz linewidth, showing good agreement.

\section*{Cavity transmission in the presence of intracavity parametric gain} 

We start from the equations of motion obtained from the hamiltonian of the system shown in Eq. (4). Assuming that we couple coherent light ($\alpha_{in}$) into the cavity at the frequency around the resonance frequency of the mode $\hat{a}_n$, the coupled mode theory equation becomes

\begin{eqnarray}
    \frac{dA_n}{dt} = -i\Delta_nA_n-ig\beta A^*_{-n}-\frac{\kappa_n}{2}A_n+\sqrt{\kappa_{n,e}}\alpha_{in}
\label{eq:eqM24}
\end{eqnarray}
\begin{eqnarray}
    \frac{dA_{-n}}{dt} = -i\Delta_{-n}A_{-n}-ig\beta A^*_{n}-\frac{\kappa_n}{2}A_{-n}
\label{eq:eqM25}
\end{eqnarray}
where $A_{\pm n}$ is the fundamental field amplitude at $\omega_{\pm n}$. We look at the steady state and get the relation between the two field amplitudes
\begin{eqnarray}
    A^*_{-n} = \frac{-ig\beta^*}{i\Delta_n-\frac{\kappa_{-n}}{2}}A_n
\label{eq:eqM26}
\end{eqnarray}
where we assumed perfect spacing of the modes in the frequency domain and thus $\Delta_n = \Delta_{-n}$. Now, we assume $\kappa_n = \kappa_{-n}$ and use the boundary condition ($\alpha_{out} = \sqrt{\kappa_{n,e}}A_n-\alpha_{in}$) to get the output field ($\alpha_{out}$) at around $\omega_n$. Finally, we get the form for the transmission in the presence of intracavity parametric gain
\begin{align}
    T &= \big\lvert \frac{\alpha_{out}}{\alpha_{in}} \big\rvert^2 \nonumber \\
    &= \Bigg\lvert 1+\frac{\kappa_{n,e}(i\Delta-\frac{\kappa_n}{2})}{\Delta^2+\frac{\kappa_n^2}{4}-g^2\lvert\beta\rvert^2}\Bigg\rvert^2.
\label{eq:eqM27}
\end{align}

This formula is used to obtain the plot in Fig S5. (b). When $g\lvert\beta\rvert=0$ (no gain), we see that the transmission is the well-known lorentzian form of the cold cavity transmission. The presence of parametric gain in the cavity has an effect of narrowing the linewidth of the mode. At a large enough gain, we may observe peak as opposed to dips due to amplification (see Fig S5.(b)). We also note that this measurement could be used to determine whether the cavity is undercoupled or overcoupled. In the undercoupled situation, the dip of the transmission would go down to zero as we increase the gain (critically coupled) and become shallower with larger gain (overcoupled). In contrast, for the overcoupled case, the dip of the transmission would not go to zero but instead would only become shallower (more overcoupled).

\setcounter{figure}{0}
\renewcommand{\figurename}{Fig S}
\begin{figure*}
  \begin{center}
      \includegraphics[width=1\textwidth]{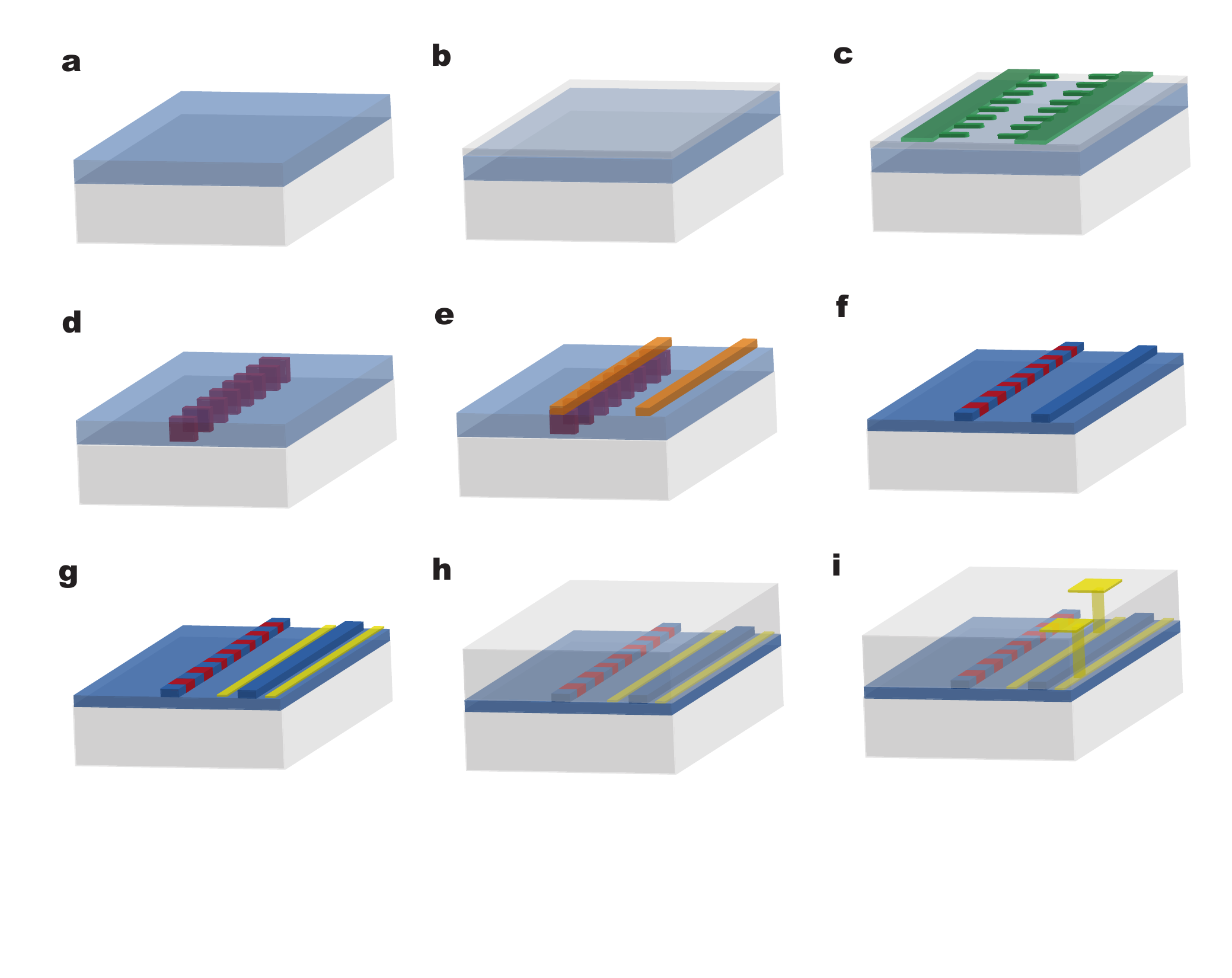}
  \end{center}
 \caption{
 \textbf{Fabrication procedure of the quantum photonic circuit.}
\textbf{a}, ~We thin down a 700-nm thin-film lithium niobate on insulator chip to get 500-nm thickness.
\textbf{b},~ Before patterning electrodes for poling, we deposit 100 nm of SiO\textsubscript{2} using a high density plasma enhanced chemical vapor deposition (HDPCVD) tool.
\textbf{c},~Using electron beam lithography and liftoff process, we fabricate chromium electrodes for poling.
\textbf{d},~We apply high voltage pulses to periodically pole the LN and then remove the electrodes by chromium etchant.
\textbf{e},~We pattern HSQ mask using electron beam lithography.
\textbf{f},~Argon ion milling the LN and acid cleaning the chip, we get patterned LN waveguides. 
\textbf{g},~Then, we pattern 100-nm gold electrodes next to the waveguides, used to controlled the phase of the transmitted light, using electron beam lithography and liftoff.
\textbf{h},~Using HDPCVD, we deposit 700 nm of SiO\textsubscript{2} for cladding.
\textbf{i},~Finally, we make vias using inductively coupled plasma etch system and gold pads with electron beam lithography and liftoff process for the contact between the external probes and the 100-nm gold electrodes adjacent to waveguides.  
}
\label{fig:Fig_S1}
\end{figure*}

\begin{figure*}
  \begin{center}
      \includegraphics[scale = 1]{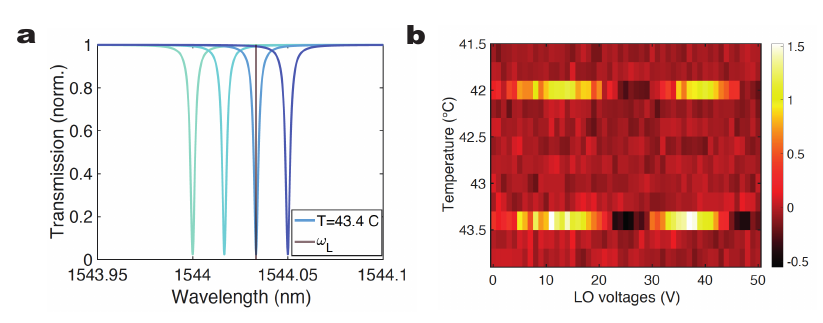}
  \end{center}
 \caption{
 \textbf{Cavity squeezing measurement with temperature stepping of the TEC setting and LO phase sweep.}
 \textbf{a}, Illustration of the mode resonance wavelength shift vs. temperature stepping of the TEC settings. Traces of different color (temperature stepping of approximately 0.4\textdegree C) indicate different temperature setting. At T = 43.4\textdegree C, the mode resonance wavelength aligns with the laser wavelength ($\omega_L$).
 \textbf{b}, Photon noise level normalized to shot noise (dB scale) vs. DC voltage applied to the local oscillator (LO) phase shifter at different temperature settings of thermoelectric cooler (TEC).
}
\label{fig:Fig_S2}
\end{figure*}

\begin{figure*}
  \begin{center}
      \includegraphics[width=1\textwidth]{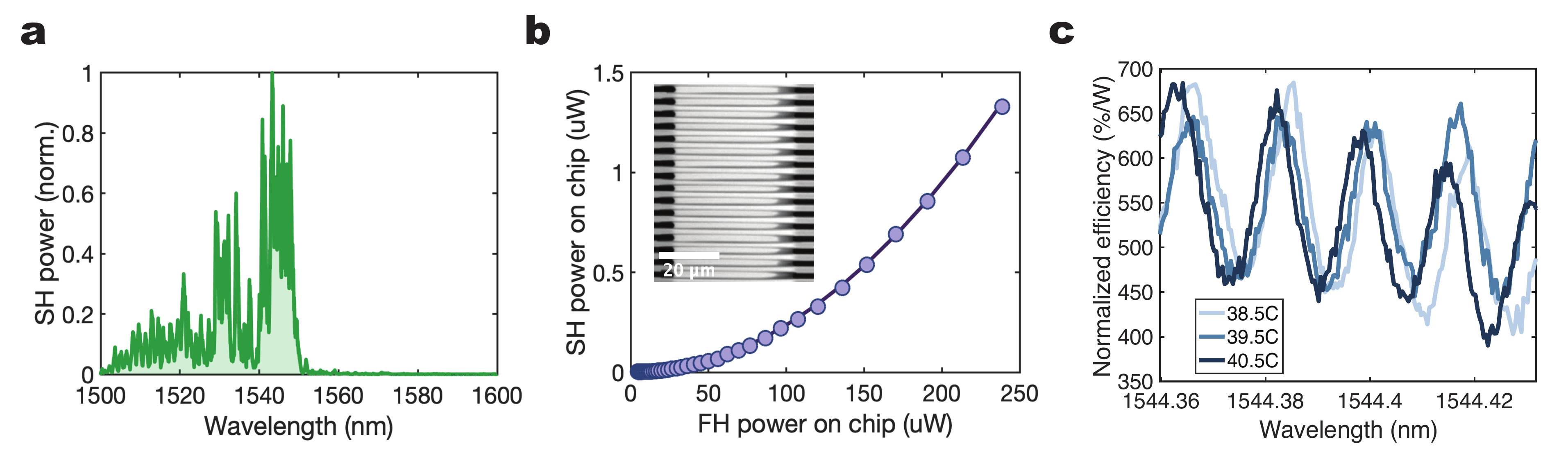}
  \end{center}
 \caption{
 \textbf{Waveguide second harmonic generator characterization. }
 \textbf{a},~SHG spectral response of the waveguide second harmonic generator. Deviation from the ideal sinc\textsuperscript{2} response is presumably due to the inhomogeneity of waveguide geometry along the 1-cm-long section.
 \textbf{b},~On-chip SH power versus on-chip FH pump power of the diagnostic SHG waveguide at the peak SH response wavelength. A quadratic fit gives normalized conversion efficiency of 1000\%/W. Inset is the second harmonic microscope image of the periodically poled section. We obtain a full-depth and uniform poling region over 30 $\mu$m in the lateral direction. 
\textbf{c},~Normalized efficiency at around the low-noise laser wavelength (1544.4 nm). Different colors indicate different temperature settings of the TEC. Oscillations with a period of approximately 20 pm originate from the Fabry-Perot cavity due to reflections at the interfaces between the waveguide and free space. The length of the cavity corresponds to the entire path length on the chip for this measurement.
}
\label{fig:Fig_S3}
\end{figure*}

\begin{figure*}
  \begin{center}
      \includegraphics[width=1\textwidth]{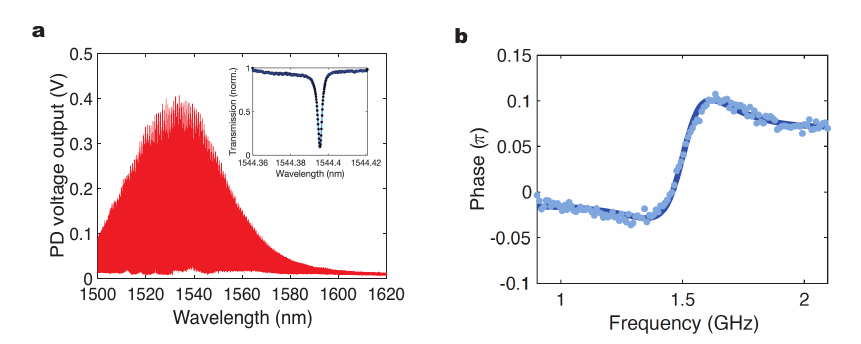}
  \end{center}
 \caption{
 \textbf{Cavity transmission profile and phase response of the mode. }
 \textbf{a},~Cavity transmission profile. The broad spectral response corresponds to the grating coupler response. Inset is the mode profile near the low-noise laser wavelength (1544.4 nm). A lorentzian fit on this mode gives \textit{Q}\textsubscript{tot} of 550\textit{k} and \textit{Q}\textsubscript{int} of 950\textit{k} assuming undercoupling.
 \textbf{b},~Phase response of a mode near 1544 nm. Light blue dots indicate data and the dark blue trace is the fit (see S.I. text).
}
\label{fig:Fig_S4}
\end{figure*}

\begin{figure*}
  \begin{center}
      \includegraphics[width=1\textwidth]{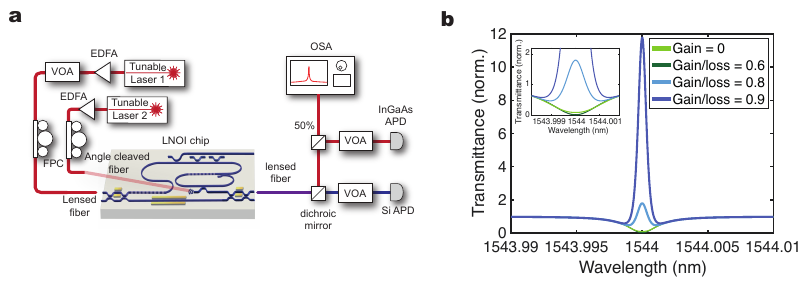}
  \end{center}
 \caption{
 \textbf{Optical parametric oscillator characterization setup and cavity transmission with gain theory. }
 \textbf{a},~Measurement setup used for characterizing the OPO. Tunable laser 1 is used for characterizing the threshold of our OPO. We use both lasers to measure the cavity transmission in the presence of parametric gain. 
 \textbf{b},~Coupled mode theory plot of cavity transmittance at different gain levels. Different colors indicate different gain/loss ratio. As the gain increase the linewidth gets narrower and at a large enough gain we observe peak due to amplification. Here, we use measured quality factors of the mode near the low-noise laser wavelength. 
}
\label{fig:Fig_S5}
\end{figure*}

\begin{figure*}
  \begin{center}
      \includegraphics[width=1\textwidth]{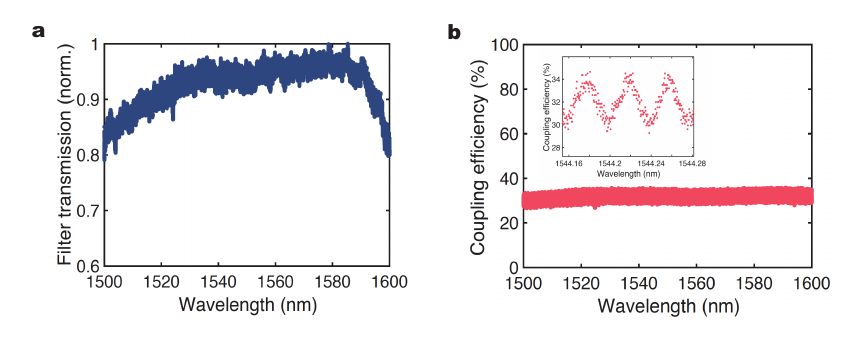}
  \end{center}
 \caption{
 \textbf{Fundamental harmonic transmission of the dichroic beamsplitter. }
 \textbf{a},~Transmission (normalized by the maximum transmission at around 1580 nm) of the dichroic beamsplitter at the fundamental harmonic verses wavelength. Assuming the maximum transmission data point has 100\% transmission, the transmission at the low-noise laser wavelength (1544.4 nm) is roughly 94$\pm$2\%. 
 \textbf{b},~Lensed fiber to waveguide coupling efficiency versus wavelength. The coupling efficiency does not show significant wavelength dependence from 1500 nm to 1600 nm.
}
\label{fig:Fig_S6}
\end{figure*}

\begin{figure*}
  \begin{center}
      \includegraphics[width=1\textwidth]{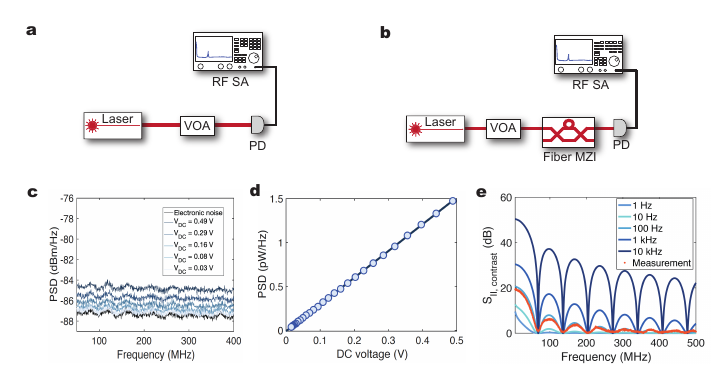}
  \end{center}
 \caption{
 \textbf{Laser noise characterization. }
  \textbf{a.}~Intensity noise characterization setup. The output of the laser goes through a variable optical attenuator (VOA) to vary the optical power incident on the photodiode (PD). The photocurrent power spectral density (PSD) from the photodetector is measured using the radio frequency spectrum analyzer (RFSA) at different incident optical powers.
 \textbf{b.}~Phase noise characterization setup. The output of the laser goes through a fiber MZI that introduces a delay (depicted as a circle on the top path) in one of the two paths. For our fiber MZI, the path length difference was 3 m, corresponding to the free spectral range (FSR) of 67 MHz. The photocurrent PSD is measured using the RFSA. \textbf{c},~Measured photocurrent power spectral density (PSD) spectrum at different optical power incident on the photodetector.
 \textbf{d},~Measured photocurrent power spectral density (PSD) averaged from 10 MHz to 500 MHz versus DC voltage read of the photodetector (electronic noise subtracted). The linear fit gives R\textsuperscript{2}~=1, showing that there is no measurable excess intensity noise.
\textbf{e},~Measured PSD spectrum after the output of the low-noise laser goes through a fiber MZI with a free spectral range of 67 MHz. Orange is the measurement data and blue traces are theory plots assuming different linewidth of the laser. The measured data aligns with the theory plot assuming the laser linewidth of 100 Hz, which is the documented value from manufacturer. 
}
\label{fig:Fig_S7}
\end{figure*}

\begin{figure}
  \begin{center}
      \includegraphics[width=1\columnwidth]{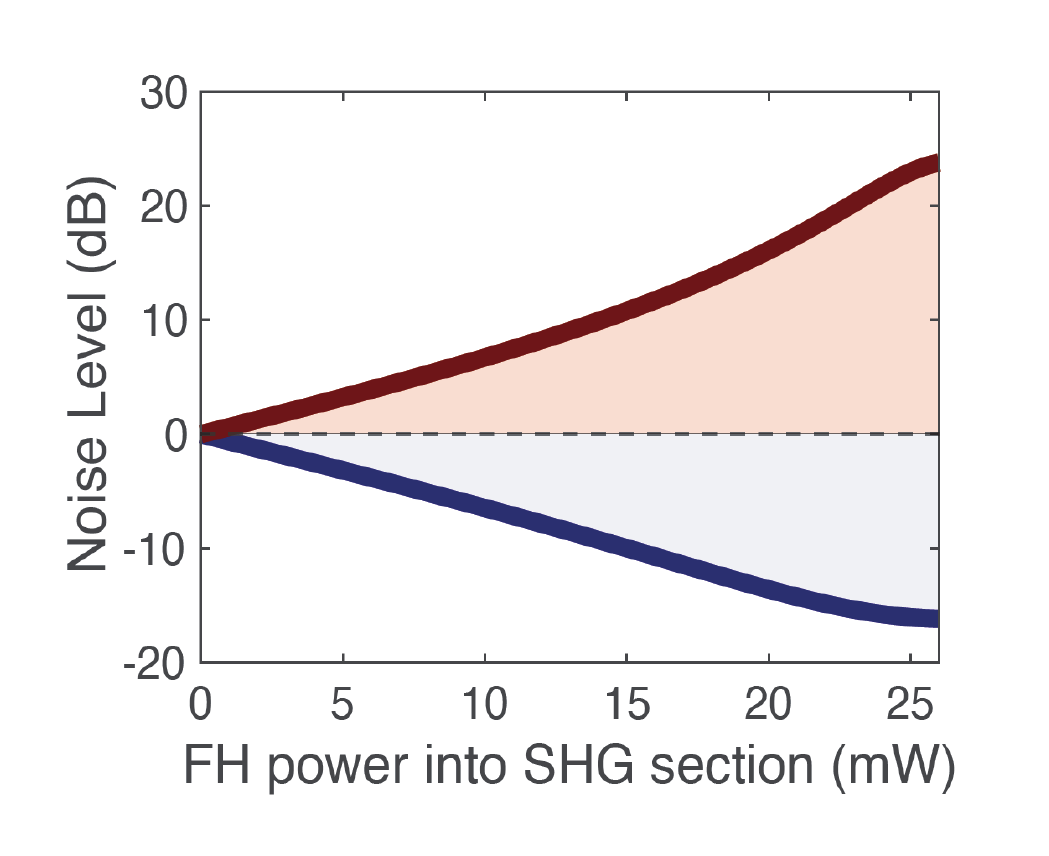}
  \end{center}
 \caption{
 \textbf{Anticipated on-chip squeezing and anti-squeezing with improved components.} Expected measured anti-squeezing (red) and squeezing (blue) versus on-chip fundamental harmonic power going into the waveguide SHG section with improved components. The length of the waveguide second harmonic generator and the OPO cavity are set to the identical values as those used in this experiment, respectively. Normalized efficiency $\eta_0$=4000\%/W-cm\textsuperscript{2} is used. For the cavity mode, we assume \textit{Q}\textsubscript{int} of 10 million and \textit{Q}\textsubscript{tot} of 200$k$.
}
\label{fig:Fig_S8}
\end{figure}

\clearpage
\bibliography{OPOSQZ}

\section*{Acknowledgements} 
This work was supported in part by the Defense Advanced Research Projects Agency (DARPA) LUMOS program (Grant No. HR0011-20-2-0046, received by A.H.S.-N.) and the DARPA Young Faculty Award (YFA, Grant No. D19AP00040, received by A.H.S.-N.). We also thank NTT Research for their financial and technical support, and the U.S. Department of Energy for their support through Grant No. DE-AC02-76SF00515, received by A.H.S.-N. In addition, the U.S. Air Force Office of Scientific Research provided a MURI grant (Grant No. FA9550-17-1-0002, received by A.H.S.-N.) that supported this research. This work was also performed at the Stanford Nano Shared Facilities (SNSF), supported by the National Science Foundation under award ECCS-2026822. We also acknowledge the Q-NEXT DOE NQI Center and the David and Lucille Packard Fellowship, and the Moore Inventor Fellowship for their support. H.S.S. acknowledges support from the Urbanek Family Fellowship. V.A. was partially supported by the Stanford Q-Farm Bloch Fellowship Program. S.G. acknowledges support from the Knut and Alice Wallenberg foundation (Grant. No. KAW 2021-0341). K.K.S.M gratefully acknowledges support from the Natural Sciences and Engineering Research Council of Canada (NSERC). A.Y.H. acknowledges NSF GRFP, Grant. No. 2146755. D.J.D. acknowledges support from the NSF GRFP (No. DGE-1656518). T.P thank Luke Qi, Nancy Yousry Ammar, and Jason Herrmann for helpful discussions.

\section*{Author contributions}
T.P. and H.S.S. designed and fabricated the device.
M.M.F. and A.H.S.-N. provided experimental and theoretical support. T.P. performed the experiments and analyzed the data with the input from H.S.S., V.A., S.G., K.K.S.M., A.Y.H., and D.J.D.. T.P., H.S.S., T.P.M., O.T.C., and F.M.M. developed the fabrication process. T.P and A.H.S.-N. wrote the manuscript.  
T.P., H.S.S, and A.H.S.-N. conceived the experiment, and A.H.S.-N. supervised all efforts. 

\section*{Competing interests}
The authors declare no competing interests.



\clearpage
\onecolumngrid

\end{document}